# Roadmap of Designing Cognitive Metrics for Explainable Artificial Intelligence (XAI)

Janet H. HSIAO
Department of Psychology, University of Hong Kong
jhsiao@hku.hk

Hilary H.T. NGAI
Department of Psychology, University of Hong Kong
hilngai@connect.hku.hk

Luyu QIU
Huawei Research, Hong Kong, China
qiuluyu@huawei.com

Yi YANG
Huawei Research, Hong Kong, China
yang.yi4@huawei.com

Caleb Chen CAO
Huawei Research, Hong Kong, China
caleb.cao@huawei.com

**Background**

Methods in cognitive science and psychology, which are traditionally used to infer cognitive processes underlying human behaviour, can be used to examine the mechanisms underlying an artificial intelligence (AI) system's behaviour and provide explanations. These methods in cognitive science and psychology include (1) measuring and comparing the system's behaviour under different conditions; (2) examining underlying factors that explain the behaviour through regression/factor analysis; and (3) building cognitive/predictive models to summarize or simulate the behaviour (i.e., using more AI; Taylor et al., 2020). Some of these ideas have been adopted in the current Explainable Artificial Intelligence (XAI) methods. For example, approaches similar to Method (1) include those using latent space traversal methods (e.g., Joshi et al., 2018; Samangouei et al., 2018), perturbation-based methods (e.g., Zhang et al., 2018), and counterfactual based methods, which examine how the AI system behaves as the input changes. Those using feature importance estimation methods, which typically use a simple model to approximate the AI system's behaviour to infer how the system uses certain features (e.g., Adebayo et al., 2018; Kim et al., 2018), and back-propagation based methods, can be categorised to be analogous to Method (2) above. The Case Based Reasoning (CBR) method that typically involves training a more interpretable CBR system, alongside a black-box AI system to provide explanations (e.g., Gates et al., 2019; Keane & Kenny, 2019), can be considered as an example of the Method (3) above.

More recently, XAI research has shifted to focus on a more pragmatic or naturalistic account of understanding, that is, whether the stakeholders understand the explanation (Páez, 2019). This point is especially important for research on evaluation methods for XAI systems. Thus, another direction where XAI research can benefit significantly from cognitive science and psychology research is ways to measure users' understanding, responses and attitudes that are relevant to evaluation of XAI systems. These measures can be used to quantify explanation quality and as feedback to the XAI system, in order to improve the explanations (e.g., Hoffman et al., 2018). The current report aims to propose suitable metrics for evaluating XAI systems from the perspective of stakeholders' cognitive states and processes.

**Measures of Cognitive States in Psychology**

Psychological measures of cognitive states and processes can be roughly categorised into subjective and objective measures. The distinctions between the two kinds of measures have long been established in studies on education (Dumont & Troelstrup, 1980), relationships (Strube & Barbour, 1983), management (Bommer et al., 1995), memory (Roediger, 1990), etc. Subjective measures refer to participants' reports of feelings and opinions (e.g., introspective self-reports, questionnaires, etc.), which are considered indirect measures of behavior. In contrast, objective measures assess the same mental construct by directly measuring participants' performance, response, or behavior on a task. Note however that the results of the two types of measures on the same cognitive construct do not always match (e.g., Argyropoulos et al., 2003; Nosek, 2007). There could be many factors underlying the misalignment. One prevalent factor is the effect of social desirability on subjective measures. People tend to give answers that are more socially desirable than their true answers to avoid negative evaluations. For example, people may attempt to maintain a positive image by exaggerating their intentions to donate in self-report measures (Sherman, 1980). Another factor contributing to the inconsistency between subjective and objective measures is that very often, mental activities are beyond one's awareness and thus difficult to measure through self-report. In contrast, objective measures are able to reveal latent changes that happen outside of our consciousness. For instance, in a phenomenon called "blindsight", patients with lesions in their primary visual cortex reported no visual experience of objects presented in their damaged visual field; however, they showed above-chance accuracy when asked to guess the location of the objects (Stoerig & Cowey, 2007).

Consistency between subjective and objective measures of a cognitive construct has shown to be task-specific. For example, in personality psychology, it has been reported that among the five dimensions of personality (including openness, conscientiousness, extraversion, agreeableness, and neuroticism), subjective and objective measures were only moderately correlated in two dimensions ($r = .25$ for Neuroticism and $r = .31$ for Extraversion; Back, Schmukle, & Egloff, 2009). Researchers typically prefer objective measures as they are direct measures of behavior/response, less influenced by task-irrelevant factors (such as social desirability), and can capture mental processes that are difficult to report or beyond awareness. Traditional objective measures directly measure observable behaviour, such as people's response accuracy, latency, or eye movements. Cognitive modelling approaches were later introduced to decompose a single behaviour measure into different components of mental processes (e.g., drift diffusion model; Ratcliff & Rouder, 1998). With the advance of technology in neuroimaging and electrophysiology, we can now also objectively measure responses in the black box of the brain and examine how they are associated with behavior in a cognitive task. Methods such as electroencephalography (EEG) and functional magnetic resonance (fMRI) are able to provide high-quality brain activation data that are precise in the temporal and spatial dimension respectively.

Accordingly, to measure stakeholders' cognitive states and processes for the purpose of evaluating the quality of an XAI system, both subjective and objective measures may be considered, and objective measures are preferred when possible.

**Cognitive States Relevant to XAI**

For the exact cognitive states or processes that are relevant to the evaluation of an XAI system, which we refer to as "cognitive metrics" below, the general approach to this issue has focused on cognitive metrics that are relevant to whether the user achieved a pragmatic understanding of the AI system, given the explanations provided by the XAI system. In the literature on cognitive science and psychology, *functional decomposition* has been one of the standard approaches to the study on cognition (Cummins, 2000). It refers to the process of explaining a cognitive capacity by breaking it down into sub-capacities. In the case of human explanation processes, it can be decomposed into how humans define, generate, select, evaluate, and communicate explanations (Miller, 2018). Based on human explanation processes, Hoffman et al. (2018) proposed a conceptual model of explanation processes of an XAI system, and how a user's pragmatic understanding of the explanations from the XAI system can be assessed at different functional stages. Specifically, in Hoffman's model, the XAI system is involved in three functional stages: explanation generation, user's mental model generation, and user's enhanced performance as the result of the mental model. Accordingly, cognitive metrics for evaluating an XAI system may be defined along these three stages.

According to Hoffman et al. (2018), at the explanation generation stage, the user's pragmatic understanding of the AI system can be assessed by the user's cognitive states or processes that reflect how good the explanation is (*Explanation Goodness*) and how satisfied the user is with the explanation (*User Satisfaction*). Between the explanation generation and the user's mental model generation stage, based on the explanations they received, users gradually generate a mental model and multiple psychological factors may modulate the process of model formation. These factors can be put into two major categories: those that are relevant to the amount of cognitive resources involved in forming a mental model, such as *User Curiosity and Attention Engagement*, and those that are relevant to the social attitude and relation with the XAI system, such as *User Trust and Reliance*. At the user's mental model generation stage, as suggested by Hoffman et al. (2018), the user's comprehension of the AI system can be assessed (*User Understanding*).

At the stage with the user's enhanced performance, the user's performance of using the AI system can be assessed, as well as its continuous change to derive a learning curve (*User Performance/Productivity*). The performance may continuously be affected by the same psychological factors that influence formation of mental models (i.e., User Curiosity and Attention Engagement, and User Trust and Reliance). It can also be continuously affected by the user's understanding that changes over time through interaction with the XAI system. With better understanding of the system, users are able to reflect on the previous explanations and gradually refine their mental models, which lead to changes in Explanation Goodness and the level of User Satisfaction with more interactions between the user and the XAI system. Thus, measuring the temporal dynamics of these cognitive metrics may be essential for evaluation purposes. In addition, as the user's performance enhancement depends on the interaction with the XAI system, measures that can assess how easy the user can correct the XAI system's actions or interact with the XAI system is another important dimension to consider (*System Controllability/Interactivity*).

Accordingly, here we propose that an XAI system can be evaluated in terms of the user's cognitive states and processes during the XAI's explanation process, along 7 cognitive metrics: *Explanation Goodness, User Satisfaction, User Curiosity/Attention Engagement, User Trust/Reliance, User Understanding, User Performance/Productivity, and System Controllability/Interaction*. Some metrics may be related to each other. The type of measures suitable for each metric may differ. Here we propose five general measurement types that may be considered in the context of the 7 cognitive metrics: Subjective measures on external stimuli, Subjective measures on internal states, Objective measures on cognitive states/processes, Subjective/objective measures based on cognitive models, and subjective/objective measures on temporal dynamics. In the table below we have summarised the recommended measure types for each cognitive metric (Table 1). We then provide more details about how we can use the recommended measures to evaluate an XAI system according to the recommended cognitive metrics.

Table 1: Recommended measure types for each cognitive metric domain, with prioritized options.

|  | Subjective: external stimuli | Subjective: internal states | Objective: Cognitive states/processes | Subjective/Objective: Cognitive models | Subjective/Objective: Temporal dynamics |
|---|---|---|---|---|---|
| Explanation Goodness | xx |  | xxx | xxx |  |
| User Satisfaction |  | xx |  | xxx | xxx |
| User Curiosity/ Attention Engagement |  | xx | xxx | xxx | xxx |
| User Trust/Reliance |  | xx | xx | xxx | xxx |
| User Understanding |  | x | xx | xxx | xxx |
| User Performance/Productivity |  |  | xxx | xxx | xxx |
| System Controllability/ Interactivity | xx |  | xxx |  |  |

xxx: Best options
xx: Good options
x: Acceptable options

**Cognitive Metrics of XAI Systems**

1) **Explanation Goodness**

Definition:
Explanation goodness is intricately linked to the goals of the user, tied together with factors such as precision and clarity. First and foremost, factors such as precision and clarity contribute to the basic "goodness" of an explanation *a priori*. Hoffman et al. (2018) refer to a list of features that are believed to innately make an explanation "good" even without context. Such features include soundness, appropriateness of detail, veridicality, usefulness, clarity, completeness, observability and dimension of variation. However, an explanation is an interactive process between two parties, and hence the "goodness" of an explanation can be considered as fairly subjective. The experience of the user in said interaction depends on the user's needs, wants and existing knowledge on the matter. The evaluation of an explanation's goodness depends on the user's subjective judgments on external stimuli — whether the explanation is good or not.

Generally, it can be assumed that "sufficient" explanations are placed at the intersection of amount of detail and level of comprehensibility. To understand what constitutes a "good" explanation can be context-dependent, with context specifically referring to the user's perception. A user may find an explanation good even though it does not satisfy all features of explanation goodness as listed above. For example, a user may prefer an explanation that is clear but incomplete to appease a bare "minimum necessary" understanding. On the other hand, a user with the goal of avoiding error may prefer a simple explanation that may even be misleading, as long as it meets an "inoculation" criterion (Miller, 2017; Miller, et al., 2017).

Important factors to consider:
Measuring explanation goodness can consider the relationship between a user's specific goal and the effectiveness of the explanation rendered during human-system interaction. Giving users different tasks with varying goals can reveal how the goals affect the goodness of evaluated explanations. For example, Vasilyeva et al. (2015) placed participants in three groups, where they had to come up with explanations with different goals — efficient, formal and functional. "Efficient" referred to assessing the causes of a certain traits in some organisms, whilst "formal" referred to categorizing said organisms into groups according to traits and "functional" referred to appraising the reason behind the emergence of such traits in the organisms. Consequently, all explanations were rated by the participants. The results reflected that participants preferred explanations that were congruent with the goal of their specific task. Moreover, the questions of the participants were affected by what the goal of their task was. The various reasons behind why the user seeks an explanation can be referred to as "triggers". These results demonstrated that the congruency between the goals and "triggers" of the user is vital in constituting a "good" explanation. Important factors to consider when measuring Explanation Goodness thus include the presence of basic explanation "goodness" features, as well as specificity of the explanation in terms of the match between the user's triggers or goals and the explanation, which may be objectively measured.

How to measure:
The goodness of an explanation can typically be evaluated using subjective measures through qualitative interviews and questionnaires with a Likert scale, with questions pertaining specifically to the Explanation Goodness Checklist (Hoffman, 2018). The design of the questionnaire should consider the explainer's goals and the specific task (Leake, 1996; Vasilyeva et al., 2017). For example, the NASA-Task Load Index (TLX) measures participants' perceived workload, and has been applied to assess a participants' attitudes towards a prototype's goodness and effectiveness (Hart & Staveland, 1988; Kulesza et al., 2010). Quantitative measures of explanation goodness can also be inferred through question-answering time, query time, accuracy, user skill level and task uncertainty as seen in the work by Lamberti and Wallace (1990). In addition, the goodness of an explanation can be objectively assessed according to whether a system can provide explanations that are relevant to the user's triggers (Brezillon & Pomerol, 1997; Mueller et al., 2019). Taking a visual classifying system as an example, an objective measurement of explanation goodness could be through the consistency between the system's output and the user's perceived visual experience.

2) **User satisfaction**

Definition:
User satisfaction refers to the subjective and internal level of fulfillment the user holds towards a provided explanation. When users receive an explanation, they may go through a process of evaluation of said explanation, through which they determine whether the explanation is satisfactory. A basic criteria to meet satisfaction is that the user understands the cause behind a phenomena per say, but cognitive biases of the user may influence the user to prefer a certain explanation (Miller 2018). While satisfaction is associated with happiness, psychologists argue that happiness is an affective construct, and satisfaction is a cognitive construct (Crooker & Near, 1997). Hence, satisfaction points towards whether the explanation fulfills a user's personal needs and wants.

Important factors to consider:
    The most significant reason for measuring user satisfaction is because satisfaction is the end-result goal in a user experience (Ferreira & de Souza Monteiro, 2020). Although it is important to note that satisfaction is a subjective feeling, for which each user may carry different meaning and benchmark, satisfaction represents the user's point of view and whether the XAI system has achieved the goal of a high quality interaction between user and system. Srinivasan and Chander (2020) have pinpointed 4 major characteristics that a system must avoid when providing explanations: too much repetition, lack of relevance, lack of coherence and excess length. An XAI system aiming for high User Satisfaction should avoid the above scenarios.

How to measure:
    User Satisfaction can be evaluated using subjective measures through a questionnaire, such as the Explanation Satisfaction Scale (Hoffman et al., 2018), which provides reasonably stable psychometric estimates of satisfaction towards a system's explanations. Interviews can also be utilized as a subjective measure of satisfaction with a more detailed recount of what contributes towards a "quality" explanation (van der Waa et al., 2021; Bilgic & Mooney, 2005). Objective measures may include measures of eye movement, heart rate variability, skin conductance, and jaw electromyography reflecting positive valenced emotional expressions associated with high User Satisfaction (Buettner, 2013; Wu et al., 2019; Yannakakis, 2008).

Relationships with other metrics:
    Explanation Goodness and User Satisfaction are intricately associated. Using a visual classifying system as an example — explaining why the output of the system is compatible and consistent with visual evidence (a potential definition of explanation goodness) could lead to enhanced satisfaction. When a model fails to provide a consistent justification for its output, it may prove unsatisfactory to users (Hendricks et al., 2016). High user satisfaction can also consequently lead to higher quality interaction between the user and system (Biran & McKeown, 2014; Donahue et al., 2013), suggesting its relationship with User Curiosity/Attention Engagement. Indeed, user satisfaction can be driven by the presence of curiosity, although they are measured differently. Naturally, if a user lacks the drive to question and seek explanations towards a system's function, the satisfaction may never be gauged.

    While Explanation Goodness and User Satisfaction are intricately associated, they are separate cognitive metrics for two reasons. Goodness evaluation concerns the user's judgments on external stimuli (whether they are good explanations or not), whereas satisfaction evaluation involves the user's internal state. Moreover, while the basic constituent of a "good" explanation can be decontextualized — *a priori*, user

satisfaction is contextualized — *a posteriori* judgment of explanations (Hoffman, 2018).

      A good measurement of user satisfaction can be thoroughly assessed through personal fulfillment towards a system's explanation. It is suggested that there is a notion of "explanation as exploration", which states that some users may find satisfaction most when exploring the system personally after experiencing an explanation (Kulesza et al., 2015; Mueller, 2019). This may prove to be a good example of assessing the user's internal drive to understand the mechanism behind a certain system, suggesting a close relationship between User Satisfaction and User Curiosity/Attention Engagement (see below).

3) **User Curiosity/Attention Engagement**

Definition
      Curiosity is a motivator for learning. It is a special form of information-seeking distinguished by the fact that it is internally driven (Kidd & Hayden, 2015). Information gap theory considers curiosity arising from the perception of a gap in knowledge and functioning like other drive states, such as hunger (Loewenstein, 1994). Building on this theory, the relationship between the amount of information and curiosity can be described with an inverted U-shaped pattern. People are least curious when there is little or excessive information, and they are most curious when there is moderate amount of information (Kang et al., 2009). Curiosity is a positive predictor of attention engagement (Schmitt & Lahroodi, 2008; Wu & Wu, 2020). Humans engage when curious, at which the desire to know causes us to set a goal and constrains our behavior (e.g., sustained attention). Attention therefore has been consistently used as an indicator of curiosity, especially for infant studies (Kidd & Hayden, 2015).

Important factors to consider
      The most significant reason for measuring curiosity in the XAI context is that the act of seeking an explanation is driven by curiosity (Hoffman et al., 2018). Studies on curiosity have ranged broadly in topic areas (e.g., play, exploration, reinforcement learning, and self-reported desire; Kidd & Hayden, 2015). It is thus helpful to consider curiosity in a situation/task specific way. Very complex information disincentivizes users as it requires excessive attention to process. In contrast, when faced with a mundane task demanding high engagement, boredom represents a failure to engage executive control networks to fully engage in said task (Danckert & Merrifield, 2018). Kulesza et al. (2013) pointed out that in a highly complex system, dynamically updating explanations may be a potential way to keep users' attention and to reach a complete understanding of the system. They also suggested the system should leverage curiosity by surprising users (e.g., generate odd outputs), and then communicate about the benefits of invested attention to important components of the explanations.

How to measure
      Hoffman et al. (2018) provided a curiosity checklist as a subjective measure and suggested that curiosity should be measured at multiple times whenever users ask for an explanation. Other subjective measures of curiosity include the Curiosity and Exploration Inventory (CEI; Kashdan et al., 2009) and the Perceptual Curiosity Scale

(PCS; Collins, Litman, & Spielberger, 2004).Users' eye-movement pattern encodes their intention of when and what to attend to as they interact with the system.

Objective measures thus can employ eye-movement data to predict/infer curiosity level. For example, via support vector machines (SVM), Hoppe et al. (2015) successfully trained a classifier to predict a participant's curiosity score from other participants' eye movement features (e.g., fixations, saccades, etc.). Their results indicate that curiosity may guide gaze behavior in a similar way across people and thus provide support for measuring curiosity with objective behavior. In recent years, machine learning methods such as hidden Markov models have become a popular choice to summarise or quantify participants eye movement behavior (e.g., the Eye Movement analysis with Hidden Markov Model, EMHMM approach; Chuk, Chan, & Hsiao, 2014; Chuk, Chan, Shimojo, & Hsiao, 2020; Hsiao, Lan, Zheng, & Chan, 2021). This modeling approach provides a more holistic measure (in contrast to individual eye movement features such as fixations and saccades) for predicting and assessing user curiosity (or other psychological attributes such as Trust; see below).

Neuroimaging techniques, such as fMRI, have identified distinguishing activity when a participant is bored versus interested. When a participant is curious, the anterior insula cortex demonstrated correlated activity with both the Default Mode Network (DMN) and the regions associated with attentional control (Danckert & Merrifield, 2018). The DMN is a set of interconnected brain regions related to spontaneous, task-unrelated, mind-wandering and lapse in attention behavior (Binder et al., 1999). A study discovered that boredom or lack of interest is characterized by rising heart rate, decreased skin conductance level, and increased cortisol levels (Merrifield & Danckert, 2014). Thus, these measures may also be used to provide objective measures of User Curiosity/Attention Engagement.

Relationships with other metrics

Good explanation may promote users' curiosity and motivate them to seek out additional information, whereas unsatisfying explanation may lead to confusion and suppress users' engagement in the system. For example, a bad explanation may either provide little information that makes users feel reticent and less engaged due to lack of knowledge, or it may be too complex or overwhelm users with redundant details. The trade-off between informativeness and simplicity of an explanation is thus important to attention engagement.

**4) User Trust/Reliance**

Definition

A widely accepted definition of trust is "a psychological state comprising the intention to accept vulnerability based upon the positive expectations of the intentions or behaviour of another" (Rousseau et al., 1998). However, trust in psychological studies has mainly focused on interpersonal trust. In the context of human-machine interaction, because people respond to technology socially, trust influences people's reliance on the machine. For example, people tend to rely on automations that they trust and reject those they do not (Lee & See, 2004). If the complexity of a system makes a complete explanation/understanding impractical, trust is important in guiding people's reliance on the system.

Important factors to consider

Trust has been examined in a multi-faceted way, including but not limited to components such as predictability, reliability, accuracy, ease of use, safety and efficiency (e.g., Adams et al., 2003; Cahour & Fourzy, 2009). Based on these theoretical components, to facilitate trust building between users and system, a good XAI should provide explanations that are predictable, reliable and of high accuracy. The same input should consistently generate the same output under a particular scenario. Trust is lost when users cannot understand the observed behaviors or decisions made by the system.

How to measure

Previous measurements of trust have largely relied on subjective scales. For example, Hoffman et al. (2018) modified the Cahour-Fourzy Scale (2009) into a Likert scale and fit the items into the context of XAI. Hoffman et al. also added new items adapted from Jian et al. (2000), the Schaefer Scale (2013) and the Madsen-Gregor Scale (2000). Overall, their scale captures four aspects of the XAI system: predictability, reliability, efficiency, and believability. Similar to curiosity, Hoffman et al. (2018) recommended that trust in the XAI context should also be measured repeatedly in a context-dependent manner.

In a study on people's perceived trustworthiness in AI-infused decision-making, Ashoori and Weisz (2019) proposed a scale with one distinct factor that is missing from Hoffman's scale, i.e., technical competence. This factor evaluates whether AI is used appropriately and correctly. Users were asked whether they think the use of an AI model is appropriate in this scenario. Trust can also be measured objectively through behavioural measures such as whether users are willing to accept advice of a machine (Shinozawa et al., 2005), follow the instruction of a machine (Gaudiello et al., 2016), and cooperate with a machine to resolve group conflicts (Jung, 2015). Users' eye movements can also be used to measure trust in a machine. For instance, based on the assumption that users who trust the system should pay less attention when the system is in automatic operation, in a study on automated vehicles, the duration of fixation on road conditions when performing a secondary task was negatively correlated with self-reported trust in the vehicles (Walker et al., 2018).

Similar to the measuring methods of User Curiosity/Attention Engagement, machine learning methods, such as hidden Markov models (e.g., EMHMM; Hsiao et al., 2021) can be used to model specific eye movement patterns related to User Trust/Reliance. EEG features (e.g., power spectrum densities and mean frequencies of each channel) and galvanic skin response (measures arousal by the skin conductivity) have also been used to predict human trust level via different classification methods (Hu et al., 2016), and thus can potentially provide objective measures of User Trust/Reliance.

Relationships with other metrics

A good XAI should avoid explanations that are difficult to understand and inefficient (i.e., low User Understanding and low Explanation Goodness), which may reduce users' reliance on the system. Explanations that can persuade users to believe in its correctness is important to generate trust (Miller, 2019). It may be more beneficial to provide a simpler but less likely explanation if it is more convincing to the users. Therefore, the goal to generate user trust can sometimes deviate from building user understanding, in a way such that persuasiveness is prioritized over knowledge transfer. The soundness (i.e., how truthful each element in an explanation is with respect to the underlying system) and completeness (i.e. the extent to which an explanation describes all of the underlying system) of an explanation jointly determines the level of trust (Kulesza, 2013). For example, oversimplified explanations are considered unsound and thus reduce the user trust, which in turn also makes users less engaged in the system.

5) **User understanding**

Definition:
User understanding refers to a user's "mental model" of a system, and it can be categorized as global, local and functional understanding. Firstly, a "mental model" borrows theories from psychology, referring to an individual's internal representation of the people, objects and environments he or she is interacting with (Norman, 1983; Rouse & Morris, 1986). Ideally, in the field of XAI, the user's mental model should build an accurate understanding of the way the system works. Basic levels of understanding should include comprehension of current states and functions, while further levels of understanding allows for prediction of future states and lower uncertainty experienced during interactions (Epley, Waytz & Cacioppo 2007; Rutjes, Willemsen & Ijsselsteijn, 2019). Naturally, explanations are required to achieve accurate mental models of complex systems. Furthermore, global understanding refers to a general comprehension of how a system works; local understanding refers to a particular understanding of a specific decision made by a system; functional understanding refers to the understanding of a system's functions and uses.

Important factors to consider:
The process of understanding is not simple nor straightforward — one must learn from an explanation, then be able to fully assimilate the provided information and lastly, present a personalized description of what has been absorbed. It can be argued that full understanding of the user may not always be required for effective XAI as it is dependent on the goals of the user (Kulesza et al., 2013). What level of understanding is sufficient for effective XAI is still an open research question, as is the philosophical definition of true understanding (Rutjes, Willemsen & Ijsselsteijn, 2019).

How to measure:
In order to gauge whether a user has understood an explanation is usually through means of regurgitation and reflection of the user's mental model of a concept (Grimm, Baumberger & Ammon, 2016; Hoffman, 2018). User understanding can also be evaluated using subjective measures through a questionnaire, such as the first question used in the Explanation Satisfaction Scale (Hoffman et al., 2018). More frequently in educational psychology, user understanding can be tested through a knowledge test objectively at different time points during learning (e.g., Zheng, Ye, & Hsiao, 2019). This concept has also been used in evaluating XAI systems, such as the grasp-ability test (Kim, 2018).
Simply put, if a user is given a topic that requires explanation, a good way of quantifying the extent of user understanding would consider the accuracy of the user's explanation. Schraagen et al. (2020) approached user understanding in a more holistic way, where different types of explanations' effects on users' understanding were evaluated at multiple points in time. Another good measurement of user understanding is that there is consistent and stable comprehension of global, local and functional systems throughout multiple points in time. Mental models can also be built according to the explanations, responses, or behavior of the user, and the similarity between such a user mental model and the model of the AI system can be used to assess the user's understanding of the AI system.

Relationships with other metrics:

The evaluations on Explanation Goodness, User Satisfaction, and User Understanding have been suggested to be related to each other. For example, Holzinger et al. (2020) proposed the System Causability Scale (SCS), which emphasizes on causability, the extent to which an explanation achieves a specified level of causal understanding. It can be evaluated in terms of effectiveness, efficiency, satisfaction related to causal understanding and its transparency for a user (Holzinger et al., 2020). User understanding has also been shown to be associated with eye movement patterns. For example, Zheng et al. (2019) showed that participants' eye movement patterns during documentary video viewing are associated with their comprehension accuracy of the video content. This result suggests that User Understanding may also be related to Curiosity/Attention Engagement.

**6) User Performance/Productivity**

Definition:

User Performance/Productivity refers to the performance or productivity of using the AI system as a result of the explanations provided by the XAI system. User performance or productivity can be measured continuously to form a learning curve. A learning curve function describes the degree of success attained during a period of instruction (i.e., score over time; Singer, 1982). It is an overarching metric that reflects the states of motivation, attention, concentration and decision making. It can also be used as a feedback tool to understand the temporal dynamics of user experience.

Important factors to consider

In cognitive psychology research, task learning performance can be influenced by multiple factors, including age (Chan, Chan, Lee, & Hsiao, 2018), learning experience (e.g., Tso, Au, & Hsiao, 2014), prior experience in other related tasks (e.g., Liu, Chuk, Yeh, & Hsiao, 2016; Hsiao, An, Zheng, & Chan, 2021), cognitive abilities (e.g., Zhang et al., 2019), cultural difference (e.g.,Thorup et al., 2018), socioemotional factors (e.g., Chan, Suen, Hsiao, Chan, & Barry, 2020; Chan, Jackson, Hsiao, Chan, & Barry, 2020), and mental health related factors (e.g., Zhang, Chan, Lau, & Hsiao, 2019). Thus, these factors should be taken into account when using users' performance to evaluate the quality of an XAI system.

In addition, potential ceiling or floor effect in the performance measure should be ruled out. Strictly speaking, a performance asymptote exists only if there is a stable value to which performance approaches arbitrarily closely as one extends training for arbitrarily many trials. In this sense, participants in many paradigms never reach the asymptote level (Gallistel, Fairhurst, & Balsam, 2004). Instead, their performance fluctuates irregularly. Such fluctuations are ubiquitous in behavioral measurements of human cognition (Gilden, Thornton, & Mallon, 1995). When learning is achieved, it is expected that people will show a gradual increase in performance (e.g., speed, accuracy) accompanied by a reduction in performance variability (Adi-Japha et al., 2008). As overarching as it is, as pointed out earlier, an individual's learning curve is subject to many factors that are not related to XAI (e.g., age, educational background, and attitudes towards technologies, etc.). It is thus impossible to set a predefined point that distinguishes the asymptote level for all users. Evaluations on the learning curve can focus on its general trend that users showing gradual improvement over time may be considered a successful implementation of the system.

How to measure

User performance of using the AI system can typically be measured through objectively measuring the user's behavior or cognitive processes/states, such as their accuracy, response times, eye movement behavior, or even brain imaging/electrophysiological (such as EEG) measures of performing a specific task using the AI system. These measures can also be modeled through cognitive or computational modeling to provide a more theory-driven description of the behavior or cognitive processes (e.g., using hidden Markov models to summarise users' eye movement behavior in the EMHMM method; Chuk et al., 2014; Chuk et al., 2020; Hsiao et al., 2021).

Learning curves also can be modelled using an equation to enable comparisons across XAI systems or users with different backgrounds. However, there is no accepted standard for a measurement of the learning curve. Research in health technologies has identified three main features of a learning curve that provide a descriptive contour of the curve. (a) the initial/starting level, capturing where the curve begins; (b) the rate of learning, capturing how quickly a particular level of learning is reached; (c) the asymptote/expert level, measuring the level at which learning stabilises (Cook et al., 2007). Other potential objective measures include number of trials/cases (or time required) for the user to learn, and rate of learning accuracy increase and variability decrease (Adi-Japha et al., 2008; Hoffman et al., 2018). Individual differences in age and knowledge state may be taken into account.

Relationship with other metrics

Overall, good explanations can facilitate the learning speed/rate, whereas bad explanations may undermine users' learning process that slow down or even prevent them to reach the expert level (a stabilised state). Similarly, high User Satisfaction, high User Curiosity/Attention Engagement, and high User Trust/Reliance are all associated with better User Performance.

7) **System Controllability / Interactivity**

Definition:

Humans are complex and social creatures, and a core component of social interaction involves the explanation of behaviors, albeit of one's own or of other's. Explanations of a system's processes provide users with a vague meaning of this black box system, and these explanations guide how people respond to, predict, and influence the system's behavior (de Graaf & Malle, 2017; Malle et al., 2000). This social interaction between user and system is the essence of linkage to how an individual or a system creates an explanation of another individual's or system's beliefs or behaviors. Controllability of the XAI system can be honed onto the dialog system in whatever context the AI aims to function for. A goal would be to develop systems that are more context-aware, flexible and tailored towards providing explanations and justifications that a specific user would want and accept.

Important factors to consider:

Pragmatic considerations, such as the knowledge state, demographic, culture and age of the individual receiving an explanation, can lead to one or another mode of explanation being preferred (Heerink et al., 2010; Prasad, 2017). For example, adults prefer to generalize on the basis of causal similarity, while younger children prefer to generalize on the basis of perceptual similarity and perform better when prompted for explanation by themselves (Sobel et al., 2007; Walker et al., 2014). Thus, how easy the users can interact and change the XAI system's behavior according to their knowledge state and goals etc., is another important factor to consider (Hoffman et al., 2018).

In addition, humans like to rely upon pre-existing knowledge of machines, minds and intelligence when interacting with AI. A measurement of a good metric reflecting on the social aspects of an AI system should be able to decrease the presence of cognitive dissonance. Cognitive dissonance is a theory proposed by Festinger (1957), who postulated that this phenomenon of cognitive conflict occurs when one experiences discomfort when encountering contradicting ideas that challenge their pre-existing mental model of a concept. Levin et al. discovered that cognitive dissonance is a factor mediating the one-way view of a human towards a system (2013). Hence, measurements of the social aspects of an AI system should strive to broaden the context of explanations and acknowledge that human-AI interaction is an interpersonal and social interchange of knowledge.

How to measure:

The extent to which individual differences, controllability, and interactivity have been taken into account in the XAI system can be assessed subjectively after the interaction through questionnaires. Such questionnaires include questions regarding the user's perception of the robot's naturalness of behavior, likability and competence (Huang & Mutlu, 2012). It can also be assessed objectively through measuring behavior such as rate of interactions between the system and the user, in order to determine whether the user increases frequency of interaction, implying improved social engagement.

Relationships with other metrics:

Intuitively, high Explanation Goodness, high User Satisfaction, high Curiosity/Attention Engagement, high User Trust/Reliance, and high User Understanding may all enhance the amount of interaction between the user and the AI system. Similarly, high System Controllability/Interactivity may also enhance quality of explanations as defined or measured using these cognitive metrics. On the other hand, while previous work has focused on making AI systems (or similarly, XAI systems) expressive and socially aware with the goals of increasing interactivity and social engagement, it may not directly improve the transparency behind the processes of a system or enhance the quality of explanations (Hoffman et al. 2014; Huang & Mutlu, 2012). Studies reveal that users typically utilize prior experience of general knowledge to determine what robots "know" (Lee et al., 2010; Wortham, Theodorou, & Bryson, 2016). This general knowledge stems from interactions with other human beings, and not interactions with other systems. When a system possesses human-like features, enhancing the social aspect of interaction, users may construct a misleading mental model of the system. This will lead to wrong understanding and inaccurate prediction of the system's actions (Epley, Waytz, & Cacioppo, 2007).

# Case Study: XAI for an Image Classification System

In this section, we will take an image classification AI system as an example, and recommend some suitable measures for its XAI system using the 7 proposed cognitive metrics.

## 1) Explanation Goodness

Subjective measurements of Explanation Goodness can be assessed through interviews and questionnaires, such as Explanation Goodness Checklist (Hoffman, 2018), Usability Questionnaire, Perceived System Accuracy (van der Waa et al., 2021), etc. As for objective measures, it can be measured as the consistency between the user's perception of the XAI system's output and the user's perceived visual experience when using the AI system. For example, a user is curious about how the AI system will classify an image containing a dog. However, the XAI system returns prototype (or counterfactual) examples of a cat's image or a dog image that differs significantly from the given dog, the Explanation Goodness of the XAI system is low. Similarly in attribution XAI methods such as salience map, if the XAI system returns a saliency map that highlights regions that differ from the user's original interested regions, the Explanation Goodness is low. Moreover, from a recent study, the intersection between salient regions highlighted and the human-annotated grounding boxes can capture whether the saliency map is compact (Li et al., 2021). Thus, a quantitative measure of the consistency from cognitive perspectives can be defined as the match in attended features (which can be measured through eye tracking, for example) when viewing the image input to the AI system for classification and when viewing the prototype examples provided by the XAI system. A higher match in attended features or regions would indicate a better explanation.

Explanation Goodness may also be reflected in users' emotional/physiological states associated with surprise (which is related to User Satisfaction; Vanhemme, 2000; Meyer et al., 1989; See below), as an explanation that is consistent with users' expectation/visual experience is typically associated with low surprise. Surprise is shown to be empirically associated with delay in simple response/involuntary focus of attention (Meyer et al., 2008), which may be captured by eye movement behavior such as a slower orienting saccade to highlighted regions on the saliency map with increased fixation duration (e.g., Horstmann & Herwig, 2015). More recently, potential electrophysiological measures (including brain activity with electroencephalogram EEG, heart rate with electrocardiography ECG, or facial expression with facial electromyography fEMG, or video analysis), especially those related to error detection and feedback processing, have also been explored (e.g., Li et al, 2018). Predictive models can be developed using these objective measures to automatically detect Explanation Goodness without relying on subjective measures. For example, a predictive model has been built using peripheral physiological signals (fEMG) to predict users' perceived relevance of text items when they are engaged in an information retrieval task (Barral et al., 2015). Another study infers the user's perceived relevance of information provided through EEG (Eugster et al., 2014) and through the influence of reading speed on pupil size (Barral, Kosunen, Jaccuci, 2014). Although such studies focus on a user's understanding of text items, this can be applied to assess a user viewing a saliency map. This is a promising direction for future research.

**2) User Satisfaction**

User Satisfaction is intrinsically an internal cognitive state, and thus is typically measured subjectively. For example, it can be assessed through interviews and questionnaires such as the Explanation Satisfaction Scale (Hoffman, 2018). User satisfaction can sometimes be inferred from behavior or emotional/physiological states. For example, when a user is satisfied with a saliency map that highlights the diagnostic features the AI system uses to classify a given dog image, the user may exhibit longer total viewing time on the highlighted region with a larger perceptual span (Buettner, 2013; Wu et al., 2019) together with positive valence that may be associated with suppression of heart rate variability (Yannakakis, 2008), as well as high galvanic skin response and high heart rate coupled with positive jaw electromyography (i.e. a positive facial expression as measured in fEMG or video analysis) (Mandryk et al., 2006; Ravaja et al., 2006). On the other hand, if a user is not satisfied with the saliency map provided, affective states may include anger and frustration, which can be objectively measured using face recognition in video analysis as well as increased electrodermal activity (EDA) (Yang & Dorneich, 2015). Thus, it is possible to build a predictive model using these measures to predict User Satisfaction (Martinez, Bengio & Yannakakis, 2013; Nasoz et al., 2003; Wu et al., 2019). Then, users' internal states can be monitored continuously and used to predict User Satisfaction as real-time feedback to the XAI system, without relying purely on subjective measures such as the use of questionnaires.

**3) User Curiosity/Attention Engagement**

Subjective measures of user curiosity and attention engagement include scales and checklists. Users can quickly check the reasons for seeking an explanation through a few clicks whenever they seek more information from the system (e.g., A Curiosity Checklist; Hoffman et al., 2018). The efficiency of the checklist enables a "quick response window" that can be administered multiple times throughout the user-AI interaction. . For example, if a user checks, "I want to know what the AI would have done if something had been different", a counterfactual XAI system may provide a good explanation. Users' trait curiosity can be measured as a baseline control when using the XAI system (e.g. The Curiosity and Exploration Inventory-II; Kashdan et al., 2009).

As for objective measures, Curiosity/Attention Engagement may be objectively measured by the frequency of the user giving the image classification AI system images to classify. It may also be inferred from behavior. Predictive models can again be built to predict subjective/objective measures of Curiosity/Attention Engagement using behavioral and physiological measures such as eye movements, EEG, and pupillometry (e.g. Fairclough, Ewing, & Roberts, 2009). Such predictive models can be used to automatically assess users' Curiosity/Attention Engagement through monitoring users' psychological/physiological states (regardless of which XAI method is used). For example, to monitor and improve user engagement, low-cost, wireless EEG was used to detect signals from FP1 regions (which is known to manage learning and concentration). Common artifacts (e.g., eye-blinks and muscle activities) were filtered out and a special two-tiered system was implemented to analyze EEG engagement levels and identify significant drops in attention in real-time (Szafir & Mutlu, 2012). The $\beta/(\alpha+\theta)$ index retrieved from cross-channel EEG data can also reflect motivation and task engagement (Xu & Zhong, 2018). High task engagement is characterized by increased beta, and decreased alpha and theta. Posture analysis such as detection of head gesture (shakes and nods), and facial expression (smiling or frown) can also infer users' feelings towards an explanation (e.g., a saliency map). Negative posture and expression can predict frustration and boredom (Kapoor et al., 2007), which are shown to be negatively correlated with curiosity and attention engagement.

**4) User Trust/Reliance**

Subjective measures of user trust and reliance divided the metric into subcomponents. Taking the questionnaire proposed by Hoffman (2018) as an example, trust was analyzed into four parts: (1) reliability, (2) predictability, (3) efficiency, and (4) believability. For each part, a sample question is provided here in the context of the prototype images selected by an XAI. (1) What is your confidence in the current prototype images in representing the whole dataset? Can you count on it to be correct all the time ? (2) Do you think the generated prototypes are predictable? (3) Do you consider the system efficient in producing prototypes? (4) Are you wary of the system generated prototypes?

Alternatively, user trust and reliance on the system can be directly shown via their behavior. Whether or not users are willing to accept and make their decisions based on the predictions/suggestions generated by the XAI is an indicator of user trust (e.g., doctors refused to make diagnosis with the ranking similarity information of the prototype patients; Wang et al., 2019). Physiological measures may be used to infer user trust during the user-AI interaction. Lower amplitude of skin conductance and higher release of oxytocin (a type of hormone that can typically be measured from blood samples) were found after the introduction of a trust-related social interaction (Kéri & Kiss, 2011). Lastly, users' brain activities may be recorded each time they see a prototype image, and these data can be analyzed to extract features that are related to trust changes. For instance, high-beta band was positively related to increased vigilance (Knyaze et al., 2002).

**5) User Understanding**

Although it is possible to directly ask users whether they understand the explanations provided by the XAI system as a subjective measure of User Understanding, the most straightforward method is to directly and objectively measure their understanding through knowledge tests. For example, the concept of the grasp-ability test to test the user's ability to answer counter-factual or propositional "X because of Y" questions has been proposed. The grasp-ability test generates questions that modify a factual prior feature (e.g. "What if the input in this image had X instead, other things equal?") and then assesses the consequences of that change in order to gauge pragmatically whether the user understood the explanation (Kim, 2018). In the case of attribution XAI methods such as salience map methods, User Understanding may be tested by asking the user to predict diagnostic features that would be used by the AI system. Moreover, since understanding is a dynamic process, there may be changes in a user's understanding throughout time.

A user's ability to predict the outcome of an image classifier has also been proposed as a measure to assess the user's understanding, as well as how transparent or explainable a system is (Lipton, 2018). Research shows that when a saliency map is present, along with the images to be classified, participants will predict the outcome of a classifier system significantly more accurately (Alqawaari, 2020). This indicates that saliency maps influence participants to saccade towards saliency features, emphasizing those features' importance.

Using objective measures such as functional magnetic resonance imaging (fMRI), user understanding can be assessed by whether external information retrieval is required while performing a task, such as predicting diagnostic features the AI system would use to categorize a certain image. Involvement of a distributed network of brain regions associated with information need and retrieval, such as the posterior cingulate, inferior parietal lobe, inferior temporal gyrus and superior frontal gyrus, can be used to predict user understanding during a grasp-ability test (Moshfeghi, Triantafillou & Pollick, 2016). In particular, evidence for the neural substrate of posterior cingulate is found to be involved with successfully recovering internal knowledge, a situation that relieves the need for an external search for information. Moreover, user understanding can also be modeled using high level behaviors, such as user dwell time, and query formulation patterns, and in low level behaviors associated with the information acquisition process such as eye movement patterns (e.g. perceptual span and fixation duration; Cole et al., 2013). It is possible that eye movement-based interaction can be used as an exemplar of a new way of user-computer interaction, in order to infer user knowledge and understanding (Jakob, 1993).

**6) User Performance/Productivity**
The most prominent feature of this metric is to provide the temporal dynamics of user performance/productivity when using an XAI system. Therefore, the metric must be measured by parameters that are time-sensitive and comparable at different time points, and since the metric is intrinsically objective, subjective measures are not applicable here. Since a learning process is affected by various factors (e.g., age, education, socioeconomic status, etc.), it is recommended that these factors were measured prior to the explanation generation as control variables. In an image classification system, shorter users' reaction latency to process the system generated prediction may be seen as an indicator of better user performance. The overtime changes in reaction latency (shorter for improved performance and vice versa) can be depicted in a learning curve format. Eye movement patterns can also reflect user performance/productivity. For example, if users demonstrate decreasing fixation duration on input features that are consistently highlighted in the previous saliency map of the same category, and decreasing fixation duration on irrelevant features, they may have learnt about the specific features that a system is sensitive to. In such cases, the efficiency of using the system is arguably improved. Similar to some other metrics, behavioral/eye movement and physiological measures can be monitored and used to build a predictive model to automatically predict users' performance as real-time feedback.

In the context of medical and health care, user performance/productivity is directly related to how clinicians and practitioners leverage the diagnosis and treatment recommended by an XAI system. For example, with the help of XAI, the average consultation time per patient (i.e., reaction time or task completion time, which is an index of user productivity; Hoffman et al., 2018) may continue to decrease whereas the diagnostic accuracy may increase over time. The two parameters

(time and accuracy) can thus be used to capture the temporal changes in doctor's performance.

**7) System Controllability / Interactivity**

In subjective measures, though there are no validated questionnaires regarding system controllability and interactivity in XAI, there are existing measures from studies that ask users to subjectively rate a system's likability, competence and naturalness (Huang & Mutlu, 2012). Other suitable topics to be explored include personalization of explanations, ease of interaction and engagement with the system, communication effectiveness and language or diction used (Rzepka & Berger, 2018). In the specific context of a saliency map method for image classifications, some questions suitable for testing the interactivity of the XAI system include connectedness to the XAI's generated saliency map example, ease of user control when choosing images for explanations, communication effectiveness of the XAI system regarding the use of saliency map for image classification and synchronicity of the XAI system's response (Yampolskiy, 2008).

When a suitable battery of questions are created, the norming process for psychological questionnaires can be carried out, in order to validate the reliability and validity of the questionnaire. The process includes pilot testing, then subsequent revisions and repeated testing until the factors of reliability and validity are fulfilled. Reliability points towards the consistency of a questionnaire's results even between different demographics; a high Cronbach's alpha estimates high internal reliability (i.e. consistency of results across items of a questionnaire) while a high Pearson's correlation indicates high test-retest reliability (i.e consistency of results over time) (Tsang et al., 2017; Cronbach & Meehl, 1955). In this context, construct validity refers to whether the questionnaire can measure the unobservable feature of controllability and interactivity. The construct validity of a questionnaire can be evaluated by estimating its association with other variables correlated with interactivity and controllability (Cohen, 1988).

As for objective measures, a system may score high in interactivity if there is increased frequency of interaction between user and system. If we assume that users may communicate with the XAI system to change the parameters used in the XAI system (e.g., the number of prototype images generated in prototype example methods, the level of importance in attribution methods, etc.), Controllability/Interactivity of the XAI system can be measured by the frequency of different types of parameter changes made by the user.

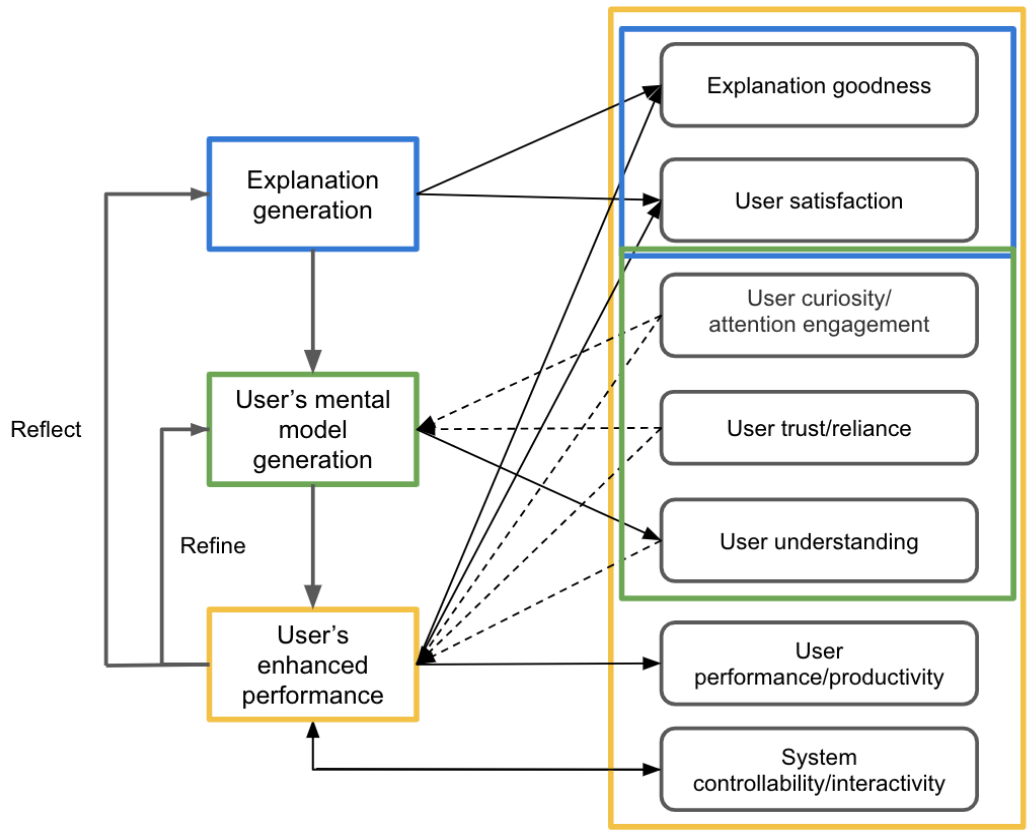

*Figure 1.* Hypothesized evaluative model for XAI, including three functional stages and seven associated cognitive metrics. The solid lines represent metrics that can be assessed at the stage, and the dashed lines represent metrics that moderate the stage.

**Acknowledgements**
This research is supported by the Computing System Theory and Technology Committee of Huawei with project number CSTT-9419577.

Zheng, Y., Ye, X., & Hsiao, J. H. (2019). Does video content facilitate or impair comprehension of documentaries? The effect of cognitive abilities and eye movement strategy. *Proceeding of the 41th Annual Conference of the Cognitive Science Society*. Austin, TX: Cognitive Science Society.